\documentclass[11pt]{article}
\usepackage{amssymb,latexsym,amsmath,amsbsy}
\usepackage{cite}
\usepackage{stmaryrd}  
\usepackage{arydshln}  
\newtheorem{theo}{Theorem}

\def\nn{\nonumber}
\def\deg{\mathop{\rm deg}\nolimits}
\def\ch{\mathop{\rm char}\nolimits}
\def\qdots{\mathinner{\mkern1mu\raise1pt\vbox{\kern7pt\hbox{.}}\mkern2mu \raise4pt\hbox{.}\mkern2mu\raise7pt\hbox{.}\mkern1mu}}
\def\Z{{\mathbb Z}}

\def\C{{\mathbb C}}
\def\gl{\mathfrak{gl}}
\def\g{\mathfrak{g}}
\def\h{\mathfrak{h}}
\def\so{\mathfrak{so}}
\def\sp{\mathfrak{sp}}
\def\osp{\mathfrak{osp}}
\def\pso{\mathfrak{pso}}
\def\lb{\llbracket}
\def\rb{\rrbracket}

\def\mybox{\hfill$\Box$}
\renewcommand{\theequation}{\arabic{section}.{\arabic{equation}}}

\begin{document}
\begin{center}
{\Large \bf
The $\Z_2\times\Z_2$-graded Lie superalgebra $\pso(2m+1|2n)$\\[2mm] and new parastatistics representations} \\[5mm]
{\bf N.I.~Stoilova}\footnote{E-mail: stoilova@inrne.bas.bg}\\[1mm] 
Institute for Nuclear Research and Nuclear Energy,\\ 
Boul.\ Tsarigradsko Chaussee 72, 1784 Sofia, Bulgaria\\[2mm] 
{\bf J.\ Van der Jeugt}\footnote{E-mail: Joris.VanderJeugt@UGent.be}\\[1mm]
Department of Applied Mathematics, Computer Science and Statistics, Ghent University,\\
Krijgslaan 281-S9, B-9000 Gent, Belgium.
\end{center}


\begin{abstract}
When the relative commutation relations between a set of $m$ parafermions and 
$n$ parabosons are of ``relative parafermion type'',
the underlying algebraic structure is the classical orthosymplectic Lie superalgebra $\osp(2m+1|2n)$.
The relative commutation relations can also be chosen differently, of ``relative paraboson type''. 
In this second case, the underlying algebraic structure is no longer an ordinary Lie superalgebra, 
but a $\Z_2\times\Z_2$-graded Lie superalgebra, denoted here by $\pso(2m+1|2n)$.
The identification of this new algebraic structure was performed by Tolstoy, amongst others.
In the present paper, we investigate the subalgebra structure of $\pso(2m+1|2n)$.
This allows us to study the parastatistics Fock spaces for this new set of $m+n$ para-operators,
as they correspond to lowest weight representations of $\pso(2m+1|2n)$.
Our main result is the construction of these Fock spaces, with a complete labeling of the basis vectors and
an explicit action of the para-operators on these basis vectors.
\end{abstract}

\vskip 10mm
\noindent Running title: $\Z_2\times\Z_2$-graded algebra $\pso(2m+1|2n)$ and parastatistics


\setcounter{equation}{0}
\section{Introduction} \label{sec:Introduction}%

The standard creation and annihilation operators of identical particles satisfy canonical commutation (boson) or
anticommutation (fermion) relations, expressed by means of commutators or anticommutators.
In 1953 Green~\cite{Green} generalized bosons to so-called parabosons and fermions to parafermions, by postulating
certain triple relations for the creation and annihilation operators, rather than just (anti)commutators.
A system of $m$ parafermion creation and annihilation operators $f_j^\pm$ ($j=1,\ldots,m$) is determined by
\begin{equation}
[[f_{ j}^{\xi}, f_{ k}^{\eta}], f_{l}^{\epsilon}]=
|\epsilon -\eta| \delta_{kl} f_{j}^{\xi} - |\epsilon -\xi| \delta_{jl}f_{k}^{\eta}, 
\label{f-rels}
\end{equation}
where $j,k,l\in \{1,2,\ldots,m\}$ and $\eta, \epsilon, \xi \in\{+,-\}$ (to be interpreted as $+1$ and $-1$
in the algebraic expressions $\epsilon -\xi$ and $\epsilon -\eta$).
Similarly, a system of $n$ pairs of parabosons $b_j^\pm$ satisfies
\begin{equation}
[\{ b_{ j}^{\xi}, b_{ k}^{\eta}\} , b_{l}^{\epsilon}]= 
(\epsilon -\xi) \delta_{jl} b_{k}^{\eta}  + (\epsilon -\eta) \delta_{kl}b_{j}^{\xi}.
\label{b-rels}
\end{equation}
These triple relations involve nested (anti)commutators, just like the Jacobi identity of Lie (super)algebras.
It was indeed shown later~\cite{Kamefuchi,Ryan} that the parafermionic algebra determined by~\eqref{f-rels} is 
the orthogonal Lie algebra $\mathfrak{so}(2m+1)$,
and that the parabosonic algebra determined by~\eqref{b-rels} is the orthosymplectic Lie superalgebra 
$\mathfrak{osp}(1|2n)$~\cite{Ganchev}. 

Having further generalization of quantum statistics in mind, Greenberg and Messiah~\cite{GM} considered combined systems
of parafermions and parabosons. 
Apart from two trivial combinations (where the parafermions and parabosons mutually commute or anticommute), 
they found two non-trivial relative commutation relations between parafermions and parabosons, also expressed by
means of triple relations. 
The first of these are the relative parafermion relations, determined by:
\begin{align}
&[[f_{ j}^{\xi}, f_{ k}^{\eta}], b_{l}^{\epsilon}]=0,\qquad [\{b_{ j}^{\xi}, b_{ k}^{\eta}\}, f_{l}^{\epsilon}]=0, \nn\\
&[[f_{ j}^{\xi}, b_{ k}^{\eta}], f_{l}^{\epsilon}]= -|\epsilon-\xi| \delta_{jl} b_k^{\eta}, \qquad
\{[f_{ j}^{\xi}, b_{ k}^{\eta}], b_{l}^{\epsilon}\}= (\epsilon-\eta) \delta_{kl} f_j^{\xi}.
\label{rel-pf}
\end{align}
The second are the so-called relative paraboson relations, to be studied in this paper.
In order to distinguish them from the parastatistics system with relative parafermion relations,
we will denote the parafermion operators by $\tilde f_j^\pm$ and the paraboson operators by $\tilde b_j^\pm$.
So the operators $\tilde f_j^\pm$ among themselves still satisfy the triple relations~\eqref{f-rels}, 
the operators $\tilde b_j^\pm$ still satisfy~\eqref{b-rels}, but the relative relations are now determined by:
\begin{align}
&[[\tilde f_{ j}^{\xi}, \tilde f_{ k}^{\eta}], \tilde b_{l}^{\epsilon}]=0,\qquad
 [\{\tilde b_{ j}^{\xi}, \tilde b_{ k}^{\eta}\}, \tilde f_{l}^{\epsilon}]=0, \nn\\
&\{\{\tilde f_{ j}^{\xi}, \tilde b_{ k}^{\eta}\}, \tilde f_{l}^{\epsilon}\}= 
|\epsilon-\xi| \delta_{jl} \tilde b_k^{\eta}, \qquad
[\{\tilde f_{ j}^{\xi}, \tilde b_{ k}^{\eta}\}, \tilde b_{l}^{\epsilon}]= (\epsilon-\eta) \delta_{kl} \tilde f_j^{\xi}.
\label{rel-pb}
\end{align}
The parastatistics algebra with relative parafermion relations, determined by~\eqref{f-rels}, 
\eqref{b-rels} and~\eqref{rel-pf}, was identified by
Palev~\cite{Palev1} and is the Lie superalgebra $\mathfrak{osp}(2m+1|2n)$.
The algebra where the relative paraboson relations hold has received attention 
in a number of papers~\cite{YJ,YJ2,KA,KK,Tolstoy2014},
but remains a somewhat intriguing structure. 
The parastatistics algebra with relative paraboson relations, determined by~\eqref{f-rels}, 
\eqref{b-rels} and~\eqref{rel-pb}, was identified as a certain $\Z_2\times\Z_2$-graded Lie superalgebra
in~\cite{YJ,YJ2,KA,Tolstoy2014}.
Tolstoy~\cite{Tolstoy2014} framed this in an even more general structure of $\Z_2\times\Z_2$-graded superalgebras.
In his notation, the parastatistics algebra with relative paraboson relations would be $\mathfrak{osp}(1,2m|2n,0)$.
In order not to confuse with the notation for real forms of Lie superalgebras, we will denote
this $\Z_2\times\Z_2$-graded Lie superalgebra here as $\pso(2m+1|2n)$.
The $\Z_2\times\Z_2$-graded Lie superalgebras are not Lie superalgebras, as will be clear in the following section.

When dealing with parastatistics, a major object of study is the Fock space.
The parafermion and paraboson Fock spaces are characterized by a parameter $p$, and their explicit construction was given recently in~\cite{parafermion} (for parafermions) and in~\cite{paraboson} (for parabosons). 
For the relative parafermion relations, the parastatistics Fock space of order $p$ was 
explicitly constructed in~\cite{SV2015}, and corresponds to an infinite-dimensional lowest weight representation $V(p)$
of the Lie superalgebra $\mathfrak{osp}(2m+1|2n)$.  
In the current paper, we shall construct the parastatistics Fock space of order $p$, for the relative paraboson relations,
as an infinite-dimensional lowest weight representation $\tilde V(p)$ of the algebra $\pso(2m+1|2n)$.  
By definition the parastatistics Fock space $\tilde V(p)$ is the Hilbert space with vacuum vector $|0\rangle$, 
defined by means of 
\begin{align}
& \langle 0|0\rangle=1, \qquad \tilde f_j^- |0\rangle = 0, \qquad \tilde b_j^- |0\rangle = 0,
\qquad (\tilde f_j^\pm)^\dagger = f_j^\mp, \qquad (\tilde b_j^\pm)^\dagger = b_j^\mp,\nn\\
& [ \tilde f_j^-,\tilde f_k^+ ] |0\rangle = p\delta_{jk}\, |0\rangle, 
\qquad \{ \tilde b_j^-,\tilde b_k^+ \} |0\rangle = p\delta_{jk}\, |0\rangle,
\label{Fock}
\end{align}
and by irreducibility under the action of the algebra $\pso(2m+1|2n)$ spanned by
the elements $\tilde f_j^\pm$, $\tilde b_j^\pm$. 
 
The structure of the paper is as follows.
In Section~\ref{sec:nosp} we recall the general definition of $\Z_2\times\Z_2$-graded Lie superalgebras.
We also argue that these structures are on the same footing as ordinary Lie algebras and Lie superalgebras,
when one is dealing with algebras defined by means of nested commutators and anticommutators.
Then the matrix form of the $\Z_2\times\Z_2$-graded Lie superalgebra $\pso(2m+1|2n)$ is given,
and the para-operators are identified.
The important observation is that $\pso(2m+1|2n)$, just like its Lie superalgebra partner $\osp(2m+1|2n)$,
contains a number of familiar subalgebras like $\so(2m+1)\oplus\sp(2n)$ and $\gl(m|n)$,
and has an appropriate Cartan subalgebra.
This allows a root space decomposition of $\pso(2m+1|2n)$,
leading to weight space decompositions of its representations.
Section~\ref{sec:C} is dealing with the important class of Fock representations $\tilde V(p)$ of $\pso(2m+1|2n)$, 
which are labeled by a positive integer~$p$.
We can take great advantage here from the fact that the relevant subalgebra structure of $\pso(2m+1|2n)$
is essentially the same as that of $\osp(2m+1|2n)$, for which the Fock representations $V(p)$
have been determined in~\cite{SV2015}.
The outcome of our detailed analysis is that the Fock representation space $\tilde V(p)$ of $\pso(2m+1|2n)$
is the same (i.e.\ has the same basis vectors) as the Fock representation space $V(p)$ of $\osp(2m+1|2n)$,
but the action of the $\pso(2m+1|2n)$ generators $\tilde f_j^\pm$, $\tilde b_k^\pm$ on these
basis vectors is a subtle ``deformation'' of the action of $f_j^\pm$, $b_k^\pm$, the generators of $\osp(2m+1|2n)$.
An instructive example is given in Section~\ref{sec:example}.
Here, we give the actions of the $\osp(3|2)$ generators and those of the $\pso(3|2)$ generators explicitly.
The actions are sufficiently simple for the reader to check by hand that the triple relations
for the two algebras are indeed of type~\eqref{rel-pf} in one case, and of type~\eqref{rel-pb} in the other case.
In the general case, the explicit actions look much more complicated.
Since the basis vectors are written in terms of $\gl(m|n)$ Gelfand-Zetlin patterns,
the matrix elements of these actions are essentially products of $\gl(m|n)$ Clebsch-Gordan coefficients
and certain reduced matrix elements.
Although these Clebsch-Gordan coefficients have been determined before~\cite{CGC, Werry},
and the reduced matrix elements coincide up to a phase with those of $\osp(2m+1|2n)$ and were given in~\cite{SV2015},
it is worth collecting all these formulas in Appendix~A, for completeness.
The paper then concludes with some remarks.

\section{$\Z_2\times\Z_2$-graded Lie superalgebras and the algebra $\pso(2m+1|2n)$}
\setcounter{equation}{0} \label{sec:nosp}

The definition of a $\Z_2\times\Z_2$-graded Lie superalgebra (LSA) was already given in~\cite{Rittenberg1,Rittenberg2}.
The physical significance of such algebraic structures has been pointed out in some papers.
Among recent works, such a $\Z_2\times\Z_2$-graded algebra appears naturally as a symmetry algebra of a differential equation called the L\'evy-Leblond equation~\cite{Aizawa1}, or in the study of $N=2$ super Schr\"odinger algebras~\cite{Aizawa2}.

As a linear space, the $\Z_2\times\Z_2$-graded LSA $\g$ is a direct sum of four graded components:
\begin{equation}
\g=\bigoplus_{\boldsymbol{a}} \g_{\boldsymbol{a}} =
\g_{(0,0)} \oplus \g_{(0,1)} \oplus \g_{(1,0)} \oplus \g_{(1,1)} 
\end{equation}
where $\boldsymbol{a}=(a_1,a_2)$ is an element of $\Z_2\times\Z_2$.
Homogeneous elements of $\g_{\boldsymbol{a}}$ are denoted by $x_{\boldsymbol{a}}, y_{\boldsymbol{a}},\ldots$,
and $\boldsymbol{a}$ is called the degree, $\deg x_{\boldsymbol{a}}$, of $x_{\boldsymbol{a}}$.
If $\g$ admits a bilinear operation (the generalized Lie bracket), denoted by $\lb\cdot,\cdot\rb$, 
satisfying the identities (grading, symmetry, Jacobi):
\begin{align}
& \lb x_{\boldsymbol{a}}, y_{\boldsymbol{b}} \rb \in \g_{\boldsymbol{a}+\boldsymbol{b}}, \label{grading}\\
& \lb x_{\boldsymbol{a}}, y_{\boldsymbol{b}} \rb = -(-1)^{\boldsymbol{a}\cdot\boldsymbol{b}} 
\lb y_{\boldsymbol{b}}, x_{\boldsymbol{a}} \rb, \label{symmetry}\\
& \lb x_{\boldsymbol{a}}, \lb y_{\boldsymbol{b}}, z_{\boldsymbol{c}}\rb \rb =
\lb \lb x_{\boldsymbol{a}}, y_{\boldsymbol{b}}\rb , z_{\boldsymbol{c}} \rb +
(-1)^{\boldsymbol{a}\cdot\boldsymbol{b}} \lb y_{\boldsymbol{b}}, \lb x_{\boldsymbol{a}}, z_{\boldsymbol{c}}\rb \rb,
\label{jacobi}
\end{align} 
where
\begin{equation}
\boldsymbol{a}+\boldsymbol{b}=(a_1+b_1,a_2+b_2)\in \Z_2\times\Z_2, \qquad
\boldsymbol{a}\cdot\boldsymbol{b} = a_1b_1+a_2b_2,
\end{equation}
then $\g$ is referred to as a $\Z_2\times\Z_2$-graded Lie superalgebra. 
By~\eqref{symmetry}, the generalized Lie bracket for homogeneous elements is either a commutator or an anticommutator.
Note that $\g_{(0,0)}$ is a Lie subalgebra, and that
$\g_{(0,0)}\oplus \g_{(0,1)}$ and $\g_{(0,0)}\oplus \g_{(1,0)}$ are Lie sub-superalgebras of the 
$\Z_2\times\Z_2$-graded Lie superalgebra $\g$. 

It is worthwhile to observe the following. When the Jacobi identity for three elements $x$, $y$ and $z$ of a Lie algebra
is expanded and rewritten by means of nested commutators or anticommutators, there are essentially only four 
ways to do this~\cite{YJ}. These are:
\begin{align*}
& [x,[y,z]]+[y,[z,x]]+[z,[x,y]]=0, \\
& [x,\{y,z\}]+[y,\{z,x\}]+[z,\{x,y\}]=0, \\
& [x,\{y,z\}]+\{y,[z,x]\}-\{z,[x,y]\}=0, \\
& [x,[y,z]]+\{y,\{z,x\}\}-\{z,\{x,y\}\}=0. 
\end{align*}
The first corresponds to the Jacobi identity for Lie algebras; the second and third appear in the 
Jacobi identity for Lie superalgebras, $\Z_2$-graded, with $x$ respectively odd or even;
the fourth expression can appear only in the Jacobi identity~\eqref{jacobi} for $\Z_2\times\Z_2$-graded Lie superalgebras.
In this sense, when one is limiting themselves to commutators and anticommutators, 
$\Z_2\times\Z_2$-graded Lie superalgebras are somehow on the same level as Lie algebras and Lie superalgebras.

Let us now fix the $\Z_2\times\Z_2$-graded Lie superalgebra $\pso(2m+1|2n)$. 
Its matrix form is closely related to, but different from that of the orthosymplectic Lie 
superalgebra $\osp(2m+1|2n)$~\cite{Kac,SV2015}.

The  $\Z_2\times\Z_2$-graded Lie superalgebra $\pso(2m+1|2n)$ can be defined as the set of all block matrices of the form
\begin{equation}
\left(\begin{array}{cc:c:cc} 
a&b&u&x&x_1 \\
c&-a^t&v&y&y_1\\ \hdashline
-v^t&-u^t&0&z&z_1\\ \hdashline
-y_1^t&-x_1^t&z_1^t&d&e\\
y^t&x^t&-z^t&f&-d^t
\end{array}\right).
\label{nosp}
\end{equation}
In~\eqref{nosp} $a$ is any $(m\times m)$-matrix, $b$ and $c$ are skew symmetric $(m\times m)$-matrices, 
$u$ and $v$ are $(m\times 1)$-matrices, $x,y,x_1,y_1$ are $(m\times n)$-matrices, $z$ and
$z_1$ are $(1\times n)$-matrices, $d$ is any $(n\times n)$-matrix,
and $e$ and $f$ are symmetric $(n\times n)$-matrices. 
The $\Z_2\times\Z_2$ grading for matrices of the form~\eqref{nosp} is schematically determined by
\begin{equation}
\left(\begin{array}{c:c:c} 
\g_{(0,0)} & \g_{(1,1)} & \g_{(0,1)} \\ \hdashline
\g_{(1,1)} & 0 & \g_{(1,0)} \\ \hdashline
\g_{(0,1)} & \g_{(1,0)} & \g_{(0,0)} 
\end{array}\right).
\label{nosp-grading}
\end{equation}
It is easy to verify that \eqref{grading}-\eqref{jacobi} are satisfied for homogeneous elements of the form~\eqref{nosp},
with the bracket given in terms of matrix multiplication:
\begin{equation}
\lb x_{\boldsymbol{a}}, y_{\boldsymbol{b}} \rb = x_{\boldsymbol{a}}\cdot y_{\boldsymbol{b}}
-(-1)^{\boldsymbol{a}\cdot\boldsymbol{b}}  y_{\boldsymbol{b}}\cdot x_{\boldsymbol{a}} .
\end{equation}

Denote by $e_{ij}$ the matrix with zeros everywhere except a $1$ on position $(i,j)$, where the row and column indices run from $1$ to $2m+2n+1$.
Just as for the $\osp(2m+1|2n)$ algebra, let us introduce the following elements: 
\begin{align}
&\tilde f_{j}^+= \sqrt{2}(e_{j, 2m+1}-e_{2m+1,j+m}), \nn\\
&\tilde f_{j}^-= \sqrt{2}(e_{2m+1,j}-e_{j+m,2m+1}); \quad (j=1,\ldots,m)
\label{f-as-e} \\
&\tilde b_{k}^+= \sqrt{2}(e_{2m+1,2m+1+n+k}+e_{2m+1+k,2m+1}), \nn\\
&\tilde b_{k}^-= \sqrt{2}(e_{2m+1,2m+1+ k}-e_{2m+1+n+k,2m+1}); \quad (k=1,\ldots,n).
\label{b-as-e}
\end{align}
Then one can check that for these elements the triple relations~\eqref{f-rels}, 
\eqref{b-rels} and~\eqref{rel-pb} are satisfied.
Moreover, the algebra is generated by these elements.
In~\cite{Tolstoy2014}, the following was proved:
\begin{theo}[Tolstoy]
The $\Z_2\times\Z_2$-graded Lie superalgebra $\g$ defined by $2m+2n$ generators $\tilde f_j^\pm$ ($j=1,\ldots,m$) 
and $\tilde b_k^\pm$ ($k=1,\ldots,n$),
where $\tilde f_j^\pm \in\g_{(1,1)}$ and $\tilde b_k^\pm \in\g_{(1,0)}$,
subject to the relations~\eqref{f-rels}, \eqref{b-rels} and~\eqref{rel-pb},
is isomorphic to $\pso(2m+1|2n)$.
\end{theo}
So for a parastatistics system with relative paraboson relations, the parafermions are of degree $(1,1)$ and
the parabosons are of degree $(1,0)$.

In the remaining part of this section, we will identify a number of relevant subalgebras of $\g=\pso(2m+1|2n)$, 
which are themselves ordinary Lie algebras or Lie superalgebras.

First, let us list the basis elements of the graded parts of $\g$ in terms of the parastatistics 
generators \eqref{f-as-e} and \eqref{b-as-e}, where the indices assume the obvious values:
\begin{align*}
& \g_{(1,1)}~:\qquad \tilde f_j^+,\quad \tilde f_j^- \\
& \g_{(1,0)}~:\qquad \tilde b_j^+,\quad \tilde b_j^- \\
& \g_{(0,0)}~:\qquad [\tilde f_j^\xi, \tilde f_k^\eta], \quad \{\tilde b_j^\xi, \tilde b_k^\eta\} \\
& \g_{(0,1)}~:\qquad \{\tilde f_j^\xi, \tilde b_k^\eta\}
\end{align*}
For all of these basis elements, the matrix form can easily be identified with the form of~\eqref{nosp}.
The elements $[\tilde f_j^\xi, \tilde f_k^\eta]$ correspond to matrices~\eqref{nosp} in which $a$, $b$ and $c$ appear,
and the other submatrices are zero. Hence the subalgebra spanned by the elements $[\tilde f_j^\xi, \tilde f_k^\eta]$
is the Lie algebra $\so(2m)$. 
From the same matrix form, it follows that the subalgebra spanned by the elements $[\tilde f_j^\xi, \tilde f_k^\eta]$ and $\tilde f_j^\pm$ is the Lie algebra $\so(2m+1)$. 
In a similar way, the elements $\{\tilde b_j^\xi, \tilde b_k^\eta\}$ correspond to matrices~\eqref{nosp} in which $d$, $e$ and $f$ appear (and others are zero), hence the subalgebra spanned by these elements is the Lie algebra $\sp(2n)$. 
So $\g_{(0,0)}= \so(2m)\oplus\sp(2n)$ and
\begin{equation}
\g_{(0,0)}\oplus\g_{(1,1)} = \so(2m+1)\oplus\sp(2n).
\end{equation}
The latter is also the even subalgebra of the Lie superalgebra $\osp(2m+1|2n)$.
Because of this, we will refer to the diagonal matrices of~\eqref{nosp}, i.e.\ the diagonal matrices of 
$\so(2m+1)\oplus\sp(2n)$, as the Cartan subalgebra $\h$ of $\pso(2m+1|2n)$. 
A basis of $\h$ is given by 
\begin{align}
& h_i=e_{ii}-e_{i+m,i+m}=-\frac12[\tilde f_i^-,\tilde f_i^+] \qquad (i=1,\ldots,m) \nn\\
& h_{m+j}=e_{2m+1+j,2m+1+j}-e_{2m+1+n+j,2m+1+n+j}=\frac12\{ \tilde b_j^-,\tilde b_j^+\} \qquad (j=1,\ldots,n). 
\label{h-basis}
\end{align}
In terms of the dual basis $\epsilon_i$ ($i=1,\ldots, m$), $\delta_j$ ($j=1,\ldots,n$) of $\h^*$, 
one has the same root space decomposition of $\pso(2m+1|2n)$ as for $\osp(2m+1|2n)$,
with the same positive and negative roots (but now graded with respect to $\Z_2\times\Z_2$ instead of $\Z_2$).
Note that $\tilde f_j^\pm$ is the root vector for $\pm\epsilon_j$, and that
$\tilde b_k^\pm$ is the root vector for $\pm\delta_k$.

An important subalgebra is identified as follows.
Let
\begin{align}
& E_{jk} = \frac12 [ \tilde f_j^+, \tilde f_k^-] \quad (j,k=1,\ldots,m),\qquad 
E_{m+j,m+k} = \frac12 \{ \tilde b_j^+, \tilde b_k^-\} \quad (j,k=1,\ldots,n); \label{evenE}\\
& E_{j,m+k} = \frac12 \{ \tilde f_j^+, \tilde b_k^-\}, \qquad E_{m+k,j} = \frac12 \{ \tilde b_k^+, \tilde f_j^-\} 
\quad (j=1,\ldots,m; k=1,\ldots,n).
\label{oddE}
\end{align}
Let us also fix a $\Z_2$-grading for these elements, which is $0$ for the elements~\eqref{evenE} and $1$ for
the elements~\eqref{oddE}.
Then it is a simple exercise to verify that the following relations hold for all elements $E_{jk}$ 
with $j,k=1,\ldots,m+n$:
\begin{equation}
E_{ij}E_{kl} - (-1)^{\deg(E_{ij})\deg(E_{kl})}E_{kl}E_{ij} =
\delta_{jk} E_{il} - (-1)^{\deg(E_{ij})\deg(E_{kl})} \delta_{il} E_{kj}.
\end{equation}
These are the relations for the standard basis elements of the Lie superalgebra $\gl(m|n)$.
So we can conclude that just like $\osp(2m+1|2n)$, also $\pso(2m+1|2n)$ contains $\gl(m|n)$ as a subalgebra.
Since the construction of Fock representations $V(p)$ of $\osp(2m+1|2n)$ in~\cite{SV2015} was based on the
subalgebra $\gl(m|n)$, with the basis vectors of $V(p)$ labeled by $\gl(m|n)$ patterns, 
the same construction can be followed for $\pso(2m+1|2n)$.
This will be done in the next section.

\section{Explicit Fock representations of $\pso(2m+1|2n)$}
\setcounter{equation}{0} \label{sec:C}

It will be convenient to have a unified notation for the parafermion and paraboson operators, being the generators 
of~$\pso(2m+1|2n)$:
\begin{equation}
\tilde c_j^\pm = \tilde f_j^\pm \quad (j=1,\ldots,m), \qquad
\tilde c_{m+j}^\pm = \tilde b_j^\pm \quad (j=1,\ldots,n),
\label{ctilde}
\end{equation}
and to use the notation $r=m+n$.
In order to construct the Fock space determined by~\eqref{Fock}, we shall make use of the subalgebras $\h$,
$\so(2m+1)\oplus\sp(2n)$ and $\gl(m|n)$ of $\pso(2m+1|2n)$ and of the analogy with $\osp(2m+1|2n)$~\cite{SV2015}.

We start again from a one-dimensional trivial $\gl(m|n)$ module $\C |0\rangle$, spanned on a vector $|0\rangle$, by setting
$\lb \tilde c_j^-,\tilde c_k^+\rb |0\rangle = p\,\delta_{jk}\, |0\rangle$, 
where $p$ is a positive integer.
By~\eqref{h-basis}, the weight of this vector in the $\epsilon$-$\delta$-basis is 
$(-\frac{p}{2},\ldots, -\frac{p}{2}|\frac{p}{2},\ldots, \frac{p}{2})$.

The subalgebra $\gl(m|n)$ of $\pso(2m+1|2n)$ can be extended to a larger subalgebra: 
\begin{equation}
{\mathfrak P} = \hbox{span} \{ \tilde c_j^-, \lb \tilde c_j^+, \tilde c_k^-\rb, \lb \tilde c_j^-, \tilde c_k^-\rb \;|\;
j,k=1,\ldots,m+n \}.
\label{P}
\end{equation}
The fact that this is actually a subalgebra follows easily from the triple relations for the $2m+2n$ generators.
One can consider ${\mathfrak P}$ as a parabolic subalgebra of $\pso(2m+1|2n)$.
By requiring $\tilde c_j^- |0\rangle =0$ ($j=1,\ldots,m+n$), the one-dimensional module 
$\C |0\rangle$ is extended to a one-dimensional ${\mathfrak P}$ module.
Just as for the Lie superalgebra $\osp(2m+1|2n)$, we can define the induced $\pso(2m+1|2n)$ 
module $\overline V(p)$ by
\begin{equation}
 \overline V(p) = \hbox{Ind}_{\mathfrak P}^{\pso(2m+1|2n)} \C|0\rangle.
 \label{defInd}
\end{equation}
This is a module for $\pso(2m+1|2n)$ with lowest weight 
$(-\frac{p}{2},\ldots, -\frac{p}{2}|\frac{p}{2},\ldots, \frac{p}{2})$.
Since the square of elements of the form $\{ \tilde f^+_j, \tilde b^+_k\}$ vanishes due to the triple relations,
\begin{align*}
\{ \tilde f^+_j, \tilde b^+_k\}^2 &= \frac 12 \{ \{ \tilde f^+_j, \tilde b^+_k\} , \{ \tilde f^+_j, \tilde b^+_k\} \} \\
& = \frac12 \{ \{ \{ \tilde f^+_j, \tilde b^+_k\} , \tilde f^+_j \}, \tilde b^+_k\} 
- \frac12 [ \tilde f^+_j, [ \{ \tilde f^+_j, \tilde b^+_k\} , \tilde b^+_k] ] = 0
\end{align*}
the Poincar\'e-Birkhoff-Witt theorem for $\pso(2m+1|2n)$ is the same as that of $\osp(2m+1|2n)$~\cite{Kac1}, 
and one can use the same basis for $\overline V(p)$ as in~\cite{SV2015}:
\begin{align}
& (\tilde c_1^+)^{k_1}\cdots (\tilde c_{m+n}^+)^{k_{m+n}} (\lb \tilde c_1^+,\tilde c_2^+\rb)^{k_{12}} 
(\lb \tilde c_1^+,\tilde c_3^+\rb)^{k_{13}} \cdots 
(\lb \tilde c_{m+n-1}^+,\tilde c_{m+n}^+\rb)^{k_{m+n-1,m+n}} |0\rangle, \label{Vpbasis}\\
& \qquad k_1,\ldots,k_{m+n},k_{12},k_{13}\ldots,k_{m-1,m},k_{m+1,m+2},k_{m+1,m+3}\ldots,k_{m+n-1,m+n} \in \Z_+, \nn\\
& \qquad k_{1,m+1},k_{1,m+2}\ldots,k_{1,m+n},k_{2,m+1},\ldots,k_{m,m+n} \in\{0,1\}. \nn
\end{align}
In general $\overline V(p)$ is not yet an irreducible module of
$\pso(2m+1|2n)$, but we have to quotient out the maximal nontrivial submodule $M(p)$ of $\overline V(p)$:
\begin{equation}
\tilde V(p) = \overline V(p) / M(p).
\label{Vp}
\end{equation}
By construction, the module $\overline V(p)$ has the same weight structure as the corresponding induced module 
for $\osp(2m+1|2n)$, hence it has the same character~\cite[(3.8)]{SV2015} in terms of $x_j=e^{\epsilon_j}$ and 
$y_k=e^{\delta_k}$:
\begin{equation}
 \ch \overline V(p) = 
(x_1\cdots x_m)^{-p/2}(y_1\cdots y_n)^{p/2} \sum_{\lambda \in {\cal{H}}} s_\lambda ({\mathbf x}|{\mathbf y}). 
\end{equation}
Herein, ${\cal{H}}$ denote the set of all partitions $\lambda$ satisfying the so-called
$(m|n)$-hook condition $\lambda_{m+1}\leq n$ \cite{Mac},
and $s_\lambda ({\mathbf x}|{\mathbf y})=s_\lambda (x_1, \ldots, x_m|y_1, \ldots, y_n)$
is the supersymmetric Schur function~\cite{Mac} 
(which is the character of a covariant $\gl(m|n)$ representation~\cite{Berele, King1990}).

Because of this structure, we can proceed as in~\cite[Section~3.1]{SV2015} in order to determine the basis vectors 
of the $\pso(2m+1|2n)$ module $\overline V(p)$.
Indeed, the decomposition with respect to $\gl(m|n)$ is the same, yielding all covariant $\gl(m|n)$ representations
labeled by a partition $\lambda\in{\cal H}$.
Thus the corresponding Gelfand-Zetlin basis (GZ) of $\gl(m|n)$~\cite{CGC} can be used, 
and the union of all these GZ-bases is then the basis for $\overline V(p)$. Let us recall
the notation of these basis vectors, as in~\cite[(3.12)]{SV2015} ($p$ is dropped from the notation of the vectors):
\begin{equation}
|\mu)\equiv |\mu)^r = \left|
\begin{array}{lclllcll}
\mu_{1r} & \cdots & \mu_{m-1,r} & \mu_{mr} & \mu_{m+1,r} & \cdots & \mu_{r-1,r}
& \mu_{rr}\\
\mu_{1,r-1} & \cdots & \mu_{m-1,r-1} & \mu_{m,r-1} & \mu_{m+1,r-1} & \cdots
& \mu_{r-1,r-1} & \\
\vdots & \vdots &\vdots &\vdots & \vdots & \qdots & & \\
\mu_{1,m+1} & \cdots & \mu_{m-1,m+1} & \mu_{m,m+1} & \mu_{m+1,m+1} & & & \\
\mu_{1m} & \cdots & \mu_{m-1,m} & \mu_{mm} & & & & \\
\mu_{1,m-1} & \cdots & \mu_{m-1,m-1} & & & & & \\
\vdots & \qdots & & & & & & \\
\mu_{11} & & & & & & &
\end{array}
\right)
= \left| \begin{array}{l} [\mu]^r \\[2mm] |\mu)^{r-1} \end{array} \right),
\label{mn}
\end{equation}
which satisfy the conditions~\cite{CGC}
\begin{equation}
 \begin{array}{rl}
1. & \mu_{ir}\in{\mathbb Z}_+ \; \hbox{are fixed and } \mu_{jr}-\mu_{j+1,r}\in{\mathbb Z}_+ , \;j\neq m,\;
 1\leq j\leq r-1,\\
 & \mu_{mr}\geq \# \{i:\mu_{ir}>0,\; m+1\leq i \leq r\};\\
2.& \mu_{is}-\mu_{i,s-1}\equiv\theta_{i,s-1}\in\{0,1\},\quad 1\leq i\leq m;\;
 m+1\leq s\leq r;\\
3. & \mu_{ms}\geq \# \{i:\mu_{is}>0,\; m+1\leq i \leq s\}, \quad m+1\leq s\leq r ;\\ 
4.& \hbox{if }\;
\mu_{m,m+1}=0, \hbox{then}\; \theta_{mm}=0; \\
5.& \mu_{is}-\mu_{i+1,s}\in{\mathbb Z}_+,\quad 1\leq i\leq m-1;\;
 m+1\leq s\leq r-1;\\
6.& \mu_{i,j+1}-\mu_{ij}\in{\mathbb Z}_+\hbox{ and }\mu_{i,j}-\mu_{i+1,j+1}\in{\mathbb Z}_+,\\
 & 1\leq i\leq j\leq m-1\hbox{ or } m+1\leq i\leq j\leq r-1.
 \end{array}
\label{cond3}
\end{equation}

Under the adjoint action in $\pso(2m+1|2n)$ of the $\gl(m|n)$ basis elements $E_{jk}$, 
\eqref{evenE}-\eqref{oddE}, the set
$(\tilde c_1^+, \tilde c_2^+,\ldots,\tilde c_{r}^+)$ is still a standard $\gl(m|n)$ tensor of rank (1,0,\ldots,0). 
So to every $\tilde c_j^+$ one can associate a unique GZ-pattern with top line
$1 0\ldots 0$:
\begin{equation}
\tilde c_j^+ \sim \begin{array}{l}1 0 \cdots 0 0 0\\[-1mm]
1 0 \cdots 0 0\\[-1mm] \cdots \\[-1mm] 0 \cdots 0\\[-1mm] \cdots\\[-1mm] 0 \end{array},
\label{cGZ}
\end{equation}
where the pattern consists of $j-1$ zero rows at the bottom, and the first $r-j+1$ rows are of the form
$1 0 \cdots 0$.
Proceeding as in~\cite{SV2015}, the tensor product rule in $\gl(m|n)$ reads
\begin{equation}
(1 0\cdots 0) \otimes ([\mu]^r) = ([\mu]^r_{+1}) \oplus ([\mu]^r_{+2}) \oplus \cdots \oplus([\mu]^r_{+r})
\label{umntensor}
\end{equation}
where $([\mu]^r) = (\mu_{1r},\mu_{2r},\ldots,\mu_{rr})$ and a subscript $\pm k$ indicates an increase of 
the $k$th label by $\pm 1$:
\begin{equation}
([\mu]^r_{\pm k}) = (\mu_{1r},\ldots,\mu_{kr}\pm 1,\ldots, \mu_{rr}).
\label{mu+k}
\end{equation}
The matrix elements of $\tilde c_j^+$ can be written as follows:
\begin{align}
(\mu' | \tilde c_j^+ | \mu ) & = 
\left( \begin{array}{ll} [\mu]^r_{+k} \\[1mm] |\mu')^{r-1} \end{array} \right| \tilde c_j^+
\left| \begin{array}{ll} [\mu]^r \\[1mm] |\mu)^{r-1} \end{array} \right) \nn\\
& = \left( 
\begin{array}{l}1 0 \cdots 0 0\\[-1mm]
1 0 \cdots 0\\[-1mm] \cdots\\[-1mm] 0 \end{array} ; 
\begin{array}{ll} [\mu]^r \\[2mm] |\mu)^{r-1} \end{array} 
\Bigg| \begin{array}{ll} [\mu]^r_{+k} \\[2mm] |\mu')^{r-1} \end{array} \right)
\times
([\mu]^r_{+k}||\tilde c^+||[\mu]^r).
\label{mmatrix}
\end{align}
In the last expression, the first factor is a $\gl(m|n)$ Clebsch-Gordan coefficient (CGC) 
determined in~\cite{CGC,Werry} and given by~\eqref{CGC1}, and the second factor is a {\em reduced matrix element}.
The possible values of the patterns $\mu'$ are determined by the $\gl(m|n)$ tensor product rule and the first line of $\mu'$
is of the form~\eqref{mu+k}. 
Just as for $\osp(2m+1|2n)$, the purpose is to find expressions for the reduced matrix elements:
\begin{equation}
\tilde G_k([\mu]^r) = \tilde G_k(\mu_{1r},\mu_{2r},\ldots,\mu_{rr}) = ([\mu]^r_{+k}||\tilde c^+||[\mu]^r),
\label{Gk}
\end{equation}
for arbitrary $r$-tuples $[\mu]^r=(\mu_{1r},\mu_{2r},\ldots,\mu_{rr})$ that correspond to highest weights.
Once these are determined, one has explicit actions of the $\pso(2m+1|2n)$ generators $\tilde c_j^\pm$ 
on a basis of $\overline V(p)$:
\begin{align}
\tilde c_j^+|\mu) & = \sum_{k,\mu'} \left( 
\begin{array}{l}1 0 \cdots 0 0\\[-1mm]
1 0 \cdots 0\\[-1mm] \cdots\\[-1mm] 0 \end{array}  ;
\begin{array}{ll} [\mu]^r \\[2mm] |\mu)^{r-1} \end{array} 
\Bigg| \begin{array}{ll} [\mu]^r_{+k} \\[2mm] |\mu')^{r-1} \end{array} \right)
\tilde G_k([\mu]^r) \left| \begin{array}{ll} [\mu]^r_{+k} \\[1mm] |\mu')^{r-1} \end{array} \right), \label{cj+r}\\
\tilde c_j^-|\mu) & = \sum_{k,\mu'} \left( 
\begin{array}{l}1 0 \cdots 0 0\\[-1mm]
1 0 \cdots 0\\[-1mm] \cdots\\[-1mm] 0 \end{array}  ;
\begin{array}{ll} [\mu]_{-k}^r \\[2mm] |\mu')^{r-1} \end{array}
\Bigg| \begin{array}{ll} [\mu]^r \\[2mm] |\mu)^{r-1} \end{array} \right)
\tilde G_k([\mu]_{-k}^r) \left| \begin{array}{ll} [\mu]^r_{-k} \\[1mm] |\mu')^{r-1} \end{array} \right). \label{cj-r}
\end{align}

For the computation of the unknown functions $\tilde G_k$, we can proceed as in~\cite{SV2015} and start
from the following action:
\begin{equation}
\{ \tilde c_r^-, \tilde c_r^+ \} |\mu) = 2h_r |\mu) = 
\Bigl(p+2(\sum_{j=1}^r \mu_{jr}-\sum_{j=1}^{r-1}\mu_{j,r-1} )\Bigr) |\mu). \label{crcr}
\end{equation}
Expresssing the left-hand side by means of~\eqref{cj+r} and~\eqref{cj-r}, 
and applying the explicit formulae for the CGCs, 
gives a complicated system of recurrence relations for the functions $\tilde G_k$.
However, since the $\gl(m|n)$ CGCs are the same, 
the relations obtained for the $\tilde G_k^2$ of $\pso(2m+1|2n)$ are the same as those obtained 
for the functions~$G_k^2$ of $\osp(2m+1|2n)$.
In other words, we must have
\begin{equation}
|\tilde G_k(\mu_{1r}, \mu_{2r},\ldots, \mu_{rr})|
=  |G_k(\mu_{1r}, \mu_{2r},\ldots, \mu_{rr})| \qquad (k=1,\ldots,r)
\label{G=G}
\end{equation}
where the expressions for $G_k([\mu]^r)\equiv G_k(\mu_{1r},\ldots ,\mu_{rr})$ have been
determined in~\cite[Proposition 4]{SV2015}, and given by~\eqref{Gkeven}-\eqref{Gm+k}.
The only thing that remains to be determined is the sign of $\tilde G_k(\mu)$.

Even before the actual determination of the sign of $\tilde G_k(\mu)$, we can determine the structure of the irreducible
representation $\tilde V(p)$.
Since $|\tilde G_k([\mu]^r)|=|G_k([\mu]^r)|$, it is zero for the same $\mu$-values.
Following the same argument as in~\cite{SV2015}, this means that all vectors $|\mu)$ with $\mu_{1r}>p$ belong to
the submodule $M(p)$, and $\tilde V(p)=\overline V(p)/M(p)$. 
In other words, the $\pso(2m+1|2n)$ module $\tilde V(p)$ has exactly the same basis vectors 
as the $\osp(2m+1|2n)$ module $V(p)$.
As a representation space, we can identify $\tilde V(p)$ and $V(p)$: $\tilde V(p)=V(p)$.
The only difference is the action of the new para-operators $\tilde f_j^\pm$, $\tilde b_k^\pm$ (generating $\pso(2m+1|2n)$)
on the basis vectors $|\mu)$ compared to the action of the old para-operators 
$f_j^\pm$, $b_k^\pm$ (generating $\osp(2m+1|2n)$) on $|\mu)$.

Our main result is the determination of these signs:
\begin{align}
& \tilde G_k([\mu]^r) = (-1)^{\mu_{1r}+\mu_{2r}+\cdots+\mu_{rr}} G_k([\mu]^r)\qquad \hbox{ for } k=1,\ldots,m, \nn\\
& \tilde G_k([\mu]^r) =  G_k([\mu]^r)\qquad \hbox{ for } k=m+1,\ldots,r.
\label{tildeG}
\end{align}
Rather than proving this directly from the relations for $\tilde G_k$, it is now more convenient to 
switch to the para-operator action, given~\eqref{cj+r}-\eqref{cj-r}.

\begin{theo}
Let $V(p)$ be the vector space with orthonormal basis vectors $|\mu)$ with $\mu_{1r}\leq p$,
and let the action of the $\osp(2m+1|2n)$ generators $f_j^\pm$ and $b_k^\pm$ be fixed and
determined~\cite{SV2015} by~\eqref{Acj+r}-\eqref{Acj-r}.
Then $V(p)$ is also an irreducible $\pso(2m+1|2n)$ module, where the action of its generators is given by:
\begin{align}
&\tilde f_j^\pm |\mu) = \pm (-1)^{\mu_{1r}+\mu_{2r}+\cdots+\mu_{rr}} f_j^\pm |\mu)\quad (j=1,\ldots,m); \nn\\
&\tilde b_k^\pm |\mu) = b_k^\pm |\mu)\quad (k=1,\ldots,n).
\label{new-action}
\end{align}
\end{theo}

\noindent {\bf Proof.} 
Now that the basis of $V(p)$ is fixed, we only need to verify that the triple relations for the
para-operators $\tilde f_j^\pm$ and $\tilde b_k^\pm$ are satisfied when acting on a basis vector $|\mu)$.
Since the action of $\tilde b_k^\pm$ is the same as that of $b_k^\pm$, the triple relations involving
only $\tilde b_k^\pm$ are automatically satisfied. 
The ones that need to be checked are those for
\[
[[\tilde f_{ j}^{\xi}, \tilde f_{ k}^{\eta}], \tilde f_{l}^{\epsilon}],\qquad
[[\tilde f_{ j}^{\xi}, \tilde f_{ k}^{\eta}], \tilde b_{l}^{\epsilon}],\qquad
[\{\tilde b_{ j}^{\xi}, \tilde b_{ k}^{\eta}\}, \tilde f_{l}^{\epsilon}], \qquad
\{\{\tilde f_{ j}^{\xi}, \tilde b_{ k}^{\eta}\}, \tilde f_{l}^{\epsilon} \} \quad\hbox{ and }\quad 
[\{\tilde f_{ j}^{\xi}, \tilde b_{ k}^{\eta}\}, \tilde b_{l}^{\epsilon}].
\]
Let us demonstrate one case, as all the others are similar.
Consider the action
\begin{equation}
\{\{\tilde f_{ j}^{\xi}, \tilde b_{ k}^{\eta}\}, \tilde f_{l}^{\epsilon} \} \;|\mu) =
\tilde f_{ j}^{\xi} \tilde b_{ k}^{\eta} \tilde f_{l}^{\epsilon} \;|\mu)
+\tilde b_{ k}^{\eta} \tilde f_{ j}^{\xi} \tilde f_{l}^{\epsilon} \;|\mu)
+\tilde f_{l}^{\epsilon} \tilde f_{ j}^{\xi} \tilde b_{ k}^{\eta} \;|\mu)
+\tilde f_{l}^{\epsilon}  \tilde b_{ k}^{\eta}\tilde f_{ j}^{\xi} \;|\mu).
\label{fbf}
\end{equation}
It is important to realize that the action of any operator $\tilde f_j^\xi$ or $\tilde b_j^\xi$ on $|\mu)$
has the effect of changing one of the top labels of this vector by $\pm 1$,
see~\eqref{cj+r}-\eqref{cj-r}, just like the actions of $f_j^\xi$ and $b_j^\xi$ on $|\mu)$. 
This means, for example, that
\begin{align*}
\tilde f_{ j}^{\xi} \tilde f_{l}^{\epsilon} \;|\mu) &=
\tilde f_{ j}^{\xi} \Bigl(\epsilon(-1)^{\mu_{1r}+\mu_{2r}+\cdots+\mu_{rr}} f_{l}^{\epsilon} \;|\mu) \Bigr) \\
&=
\xi (-1)^{\mu_{1r}+\mu_{2r}+\cdots+\mu_{rr}\pm 1} f_{ j}^{\xi} 
\Bigl(\epsilon (-1)^{\mu_{1r}+\mu_{2r}+\cdots+\mu_{rr}} f_{l}^{\epsilon} \;|\mu) \Bigr) =
- \xi\epsilon f_{ j}^{\xi} f_{l}^{\epsilon} \;|\mu) .
\end{align*}
As a consequence, one can write:
\begin{align}
\{\{\tilde f_{ j}^{\xi}, \tilde b_{ k}^{\eta}\}, \tilde f_{l}^{\epsilon} \} \;|\mu) 
&=
\tilde f_{ j}^{\xi} \tilde b_{ k}^{\eta} \tilde f_{l}^{\epsilon} \;|\mu)
+\tilde b_{ k}^{\eta} \tilde f_{ j}^{\xi} \tilde f_{l}^{\epsilon} \;|\mu)
+\tilde f_{l}^{\epsilon} \tilde f_{ j}^{\xi} \tilde b_{ k}^{\eta} \;|\mu)
+\tilde f_{l}^{\epsilon}  \tilde b_{ k}^{\eta}\tilde f_{ j}^{\xi} \;|\mu) \nn\\
&=  \xi\epsilon \Bigl( f_{ j}^{\xi}  b_{ k}^{\eta}  f_{l}^{\epsilon} \;|\mu)
- b_{ k}^{\eta}  f_{ j}^{\xi}  f_{l}^{\epsilon} \;|\mu)
- f_{l}^{\epsilon}  f_{ j}^{\xi} b_{ k}^{\eta} \;|\mu)
+ f_{l}^{\epsilon}  b_{ k}^{\eta} f_{ j}^{\xi} \;|\mu) \Bigr)
\nn \\
&= \xi\epsilon\ [[ f_{ j}^{\xi}, b_{ k}^{\eta}], f_{l}^{\epsilon} ] \;|\mu).
\end{align}
In the right hand side we can use the triple relation~\eqref{rel-pf} for the $\osp(2m+1|2n)$ generators, yielding
\begin{equation}
\{\{\tilde f_{ j}^{\xi}, \tilde b_{ k}^{\eta}\}, \tilde f_{l}^{\epsilon} \} \;|\mu) =
- \xi\epsilon |\epsilon-\xi| \delta_{jl} b_k^\eta \;|\mu).
\label{fbf2}
\end{equation}
But since $\epsilon$ and $\xi$ take values in $\{-1,+1\}$ only, one has $- \xi\epsilon |\epsilon-\xi|=|\epsilon-\xi|$.
Thus~\eqref{fbf2} leads to
\begin{equation}
\{\{\tilde f_{ j}^{\xi}, \tilde b_{ k}^{\eta}\}, \tilde f_{l}^{\epsilon} \} \;|\mu) 
= |\epsilon-\xi| \delta_{jl} b_k^\eta \;|\mu) \\
= |\epsilon-\xi| \delta_{jl} \tilde b_k^\eta \;|\mu).
\label{fbf3}
\end{equation}
This proves that the triple relation~\eqref{rel-pb} for 
$\{\{\tilde f_{ j}^{\xi}, \tilde b_{ k}^{\eta}\}, \tilde f_{l}^{\epsilon} \}$ holds,
when acting on a basis vector $|\mu)$.\mybox

To conclude, we have shown that the Fock representation spaces of $\pso(2m+1|2n)$ are the same
as those of $\osp(2m+1|2n)$.
The difference comes from a phase factor, determined in~\eqref{new-action}.
By this result, the Fock representations of the $\Z_2\times\Z_2$-graded LSA $\pso(2m+1|2n)$ are completely understood.

\setcounter{equation}{0}
\section{Example} \label{sec:example}

To illustrate the Fock representations of the algebraic structures given in general in the previous sections,
let us give the simple example for $m=n=1$. 
In this case there is just one parafermion and one paraboson creation and annihilation operator.
When these satisfy the relative parafermion relations, they generate the Lie superalgebra $\osp(3|2)$; 
when they satisfy the relative paraboson relations, they generate the $\Z_2\times\Z_2$-graded Lie superalgebra $\pso(3|2)$.

For a positive integer $p$, the Fock representation space $V(p)$ of $\osp(3|2)$ and $\pso(3|2)$ is the same, 
but the action of the para-operators is different.
Let us first describe the basis vectors of $V(p)$.
This basis is given by all vectors
\[
|\mu)= \left| \begin{array}{l} \mu_{12}, \mu_{22} \\ \mu_{11} \end{array} \right), \qquad
\mu_{12},\mu_{22},\mu_{11} \in\Z_{\geq 0}=\{0,1,2,\ldots\}
\]
where $\mu_{11} \in\{\mu_{12}, \mu_{12}-1 \}$, $\mu_{12}\leq p$, and
\[
\hbox{ if } \mu_{22}>0 \hbox{ then } \mu_{12}>0.
\]

For the Lie superalgebra $\osp(3|2)$, the parafermion and paraboson operators are denoted by $f^\pm$ and $b^\pm$, 
and their action on the above vectors is:
\begin{align}
f^+ \left| \begin{array}{l} \mu_{12}, \mu_{22} \\ \mu_{12} \end{array} \right) 
& = {G}_1(\mu_{12},\mu_{22}) 
\left| \begin{array}{l} \mu_{12}+1, \mu_{22} \\ \mu_{12}+1 \end{array} \right)
 ,\label{fp0}\\
f^+ \left| \begin{array}{l} \mu_{12}, \mu_{22} \\ \mu_{12}-1 \end{array} \right) 
& = \sqrt{\frac{\mu_{12}+\mu_{22}}{\mu_{12}+\mu_{22}+1}} {G}_1(\mu_{12},\mu_{22}) 
\left| \begin{array}{l} \mu_{12}+1, \mu_{22} \\ \mu_{12} \end{array} \right) \nn\\
 &-\sqrt{\frac{1}{\mu_{12}+\mu_{22}+1}} {G}_2(\mu_{12},\mu_{22}) 
\left| \begin{array}{l} \mu_{12},\mu_{22}+1 \\ \mu_{12} \end{array} \right),\label{fp1}\\
b^+ \left| \begin{array}{l} \mu_{12}, \mu_{22} \\ \mu_{12} \end{array} \right) 
& = \sqrt{\frac{1}{\mu_{12}+\mu_{22}+1}} G_1(\mu_{12},\mu_{22}) 
\left| \begin{array}{l} \mu_{12}+1, \mu_{22} \\ \mu_{12} \end{array} \right)\nn\\
 &+\sqrt{\frac{\mu_{12}+\mu_{22}}{\mu_{12}+\mu_{22}+1}} G_2(\mu_{12},\mu_{22}) 
\left| \begin{array}{l} \mu_{12},\mu_{22}+1 \\ \mu_{12} \end{array} \right),\label{bp0}\\
b^+ \left| \begin{array}{l} \mu_{12}, \mu_{22} \\ \mu_{12}-1 \end{array} \right) 
& = -G_2(\mu_{12},\mu_{22}) 
\left| \begin{array}{l} \mu_{12}, \mu_{22}+1 \\ \mu_{12}-1 \end{array} \right)
 ,\label{bp1} \\
f^- \left| \begin{array}{l} \mu_{12}, \mu_{22} \\ \mu_{12} \end{array} \right) 
& = {G}_1(\mu_{12}-1, \mu_{22}) 
\left| \begin{array}{l} \mu_{12}-1, \mu_{22} \\ \mu_{12}-1 \end{array} \right)\nn\\
&-\sqrt{\frac{1}{\mu_{12}+\mu_{22}}} {G}_2(\mu_{12}, \mu_{22}-1) 
\left| \begin{array}{l} \mu_{12},\mu_{22}-1 \\ \mu_{12}-1 \end{array} \right) ,\label{fm0}\\
f^- \left| \begin{array}{l} \mu_{12}, \mu_{22} \\ \mu_{12}-1 \end{array} \right) 
& = \sqrt{\frac{\mu_{12}+\mu_{22}-1}{\mu_{12}+\mu_{22}}} {G}_1(\mu_{12}-1, \mu_{22}) 
\left| \begin{array}{l} \mu_{12}-1, \mu_{22} \\ \mu_{12}-2 \end{array} \right) 
 ,\label{fm1}\\
b^- \left| \begin{array}{l} \mu_{12}, \mu_{22} \\ \mu_{12} \end{array} \right) 
& = \sqrt{\frac{\mu_{12}+\mu_{22}-1}{\mu_{12}+\mu_{22}}} G_2(\mu_{12}, \mu_{22}-1) 
\left| \begin{array}{l} \mu_{12},\mu_{22}-1 \\ \mu_{12} \end{array} \right),\label{bm0}\\
b^- \left| \begin{array}{l} \mu_{12}, \mu_{22} \\ \mu_{12}-1 \end{array} \right) 
&=\sqrt{\frac{1}{\mu_{12}+\mu_{22}}} G_1(\mu_{12}-1, \mu_{22}) 
\left| \begin{array}{l} \mu_{12}-1, \mu_{22} \\ \mu_{12}-1 \end{array} \right)\nn\\
& - G_2(\mu_{12},\mu_{22}-1) 
\left| \begin{array}{l} \mu_{12}, \mu_{22}-1 \\ \mu_{12}-1 \end{array} \right)
 . \label{bm1}
\end{align}
If a vector in the right hand side of such actions does not belong to $V(p)$, 
the corresponding term should be deleted.
The functions $G_1$ and $G_2$ are given by:
\begin{align}
G_1(\mu_{12},\mu_{22}) & = \sqrt{ 
\frac{\mu_{12}(\mu_{12}+\mu_{22}+1)(p-\mu_{12})}{\mu_{12}+\mu_{22}}}, \qquad\hbox{if $\mu_{22}$ is even} 
\label{F1even}\\
G_1(\mu_{12},\mu_{22}) & = \sqrt{ 
\mu_{12}(p-\mu_{12})}, \qquad \hbox{if $\mu_{22}$ is odd} \label{F1odd}\\
G_2(\mu_{12},\mu_{22}) & = \sqrt{ 
\mu_{12}+\mu_{22}+1}, \qquad \hbox{if $\mu_{22}$ is even} \label{F2even}\\
G_2(\mu_{12},\mu_{22}) & = \sqrt{ 
\frac{(\mu_{22}+1)(p+\mu_{22}+1)}{\mu_{12}+\mu_{22}}}, \qquad \hbox{if $\mu_{22}$ is odd.} \label{F2odd}
\end{align}

For the $\Z_2\times\Z_2$ Lie superalgebra $\pso(3|2)$, the parafermion and paraboson operators are denoted by $\tilde f^\pm$ and $\tilde b^\pm$, and their action on the above vectors can simply be described by:
\begin{equation}
\tilde f^\pm |\mu) = \pm (-1)^{\mu_{12}+\mu_{22}} f^\pm |\mu), \qquad
\tilde b^\pm |\mu) = b^\pm |\mu).
\end{equation}
It is now not difficult to check the triple relations explicitly by hand, since the actions are fairly simple.
For example, one finds for the $\osp(3|2)$ generators that
\[
[[f^+,b^\pm],f^-]\; |\mu) = -2 b^\pm\;|\mu),
\]
whereas for the $\pso(3|2)$ generators:
\[
\{ \{ \tilde f^+,\tilde b^\pm\},\tilde f^-\}\; |\mu) = 2 \tilde b^\pm\;|\mu).
\]

\setcounter{equation}{0}
\section{Conclusion and remarks} \label{sec:summary}

As emphasized in~\cite{YJ,YJ2}, the description of a combined system of paraboson and parafermion
operators (parastatistics operators) generalizes two notions.
On the one hand, it generalizes particle statistics and remains to be relevant in theoretical physics.
On the other hand, it unifies Bose and Fermi statistics, the basic feature of supersymmetric theories.

Since the introduction of parabosons and parafermions~\cite{Green} and their unification~\cite{GM},
it is known that there are essentially two non-trivial ways of combining them.
For the first way, the relative parafermion relations, it took almost 20 years to 
recognize these operators as generators of an orthosymplectic Lie superalgebra~\cite{Palev1}.
For the second way, the relative paraboson relations, it took another 20 extra years to
recognize them as the generators of a particular $\Z_2\times\Z_2$-graded superalgebra.

The identification of a set of para-operators with a known algebraic structure is important,
not only because it helps to understand the nested (anti)commutation relations that determine the operators,
but especially because it helps to comprehend and construct the Fock space representations.
Such Fock space representations are far from trivial.
For a set of $m$ parafermions and $n$ parabosons with relative parafermion relations,
the Fock spaces are certain infinite-dimensional lowest weight representations of the Lie
superalgebra $\osp(2m+1|2m)$, the lowest weight vector corresponding to the vacuum state.
Their explicit construction was completed only recently~\cite{SV2015}.
The basis states of these Fock spaces can no longer simply be expressed as creation operators 
acting on the vacuum.
Only by introducing $\gl(m|n)$ Gelfand-Zetlin patterns for the basis vectors of these representations, 
one could develop techniques to compute explicitly the action of creation and annihilation operators
(of the para-operators) on these basis vectors, and thus resolve the structure of the Fock spaces~\cite{SV2015}.

For a set of $m$ parafermions and $n$ parabosons with relative paraboson relations,
we have shown that the Fock spaces are certain infinite-dimensional lowest weight representations 
of the new $\Z_2\times\Z_2$-graded Lie superalgebra $\pso(2m+1|2m)$.
Our main work in this paper was to unravel the subalgebra structure of $\pso(2m+1|2m)$ and 
its root space decomposition.
This enabled us to deduce that the same $\gl(m|n)$ Gelfand-Zetlin patterns as for $\osp(2m+1|2n)$ representations 
can be used to label the basis vectors of the $\pso(2m+1|2n)$ Fock representations.
Having the experience for $\osp(2m+1|2n)$, we could easily extend our technique and 
compute the explicit actions of the new para-operators on these basis vectors of the 
$\pso(2m+1|2n)$ Fock spaces.

It turned out that the action of the new para-operators (i.e. with relative paraboson relations)
is related to that of the old para-operators (i.e. with relative parafermion relations)
by means of certain phase factors.
Although this fairly simple relation holds between Fock spaces of $\osp(2m+1|2n)$ and those of $\pso(2m+1|2n)$,
it is far from clear that this relation would extend to 
all irreducible representations of this $\Z_2\times\Z_2$-graded Lie superalgebra.
Nevertheless, our analysis is a great step forward in understanding these new $\Z_2\times\Z_2$-graded 
algebras and their representations, and solves some of the mystery that was associated with the new parastatistics algebra.

\appendix 
\section{Appendix}
\renewcommand{\theequation}{\Alph{section}.\arabic{equation}}
\setcounter{equation}{0}

In order for this paper to be self-consistent, we shall recall in this Appendix the explicit actions 
of the $\osp(2m+1|2n)$ generators on the basis vectors $|\mu)$ of $V(p)$, obtained in~\cite{SV2015}.
Another reason is that this allows us to correct some minor misprints in the earlier published 
expressions for the related reduced matrix elements~\cite{SV2015} and Clebsch-Gordan coefficients~\cite{CGC}.

The notation for the basis vectors has been described in~\eqref{mn}-\eqref{cond3}.
The notation for the $\osp(2m+1|2n)$ parafermion and paraboson operators is
\begin{equation}
c_j^\pm = f_j^\pm \quad (j=1,\ldots,m), \qquad
c_{m+j}^\pm = b_j^\pm \quad (j=1,\ldots,n),
\label{c}
\end{equation}
The explicit action is given by:
\begin{align}
 c_j^+|\mu) & = \sum_{k,\mu'} \left( 
\begin{array}{l}1 0 \cdots 0 0\\[-1mm]
1 0 \cdots 0\\[-1mm] \cdots\\[-1mm] 0 \end{array}  ;
\begin{array}{ll} [\mu]^r \\[2mm] |\mu)^{r-1} \end{array} 
\Bigg| \begin{array}{ll} [\mu]^r_{+k} \\[2mm] |\mu')^{r-1} \end{array} \right)
 G_k([\mu]^r) \left| \begin{array}{ll} [\mu]^r_{+k} \\[1mm] |\mu')^{r-1} \end{array} \right), \label{Acj+r}\\
 c_j^-|\mu) & = \sum_{k,\mu'} \left( 
\begin{array}{l}1 0 \cdots 0 0\\[-1mm]
1 0 \cdots 0\\[-1mm] \cdots\\[-1mm] 0 \end{array}  ;
\begin{array}{ll} [\mu]_{-k}^r \\[2mm] |\mu')^{r-1} \end{array}
\Bigg| \begin{array}{ll} [\mu]^r \\[2mm] |\mu)^{r-1} \end{array} \right)
 G_k([\mu]_{-k}^r) \left| \begin{array}{ll} [\mu]^r_{-k} \\[1mm] |\mu')^{r-1} \end{array} \right). \label{Acj-r}
\end{align}
Herein, the pattern 
\[
\begin{array}{l}1 0 \cdots 0 0\\[-1mm]
1 0 \cdots 0\\[-1mm] \cdots\\[-1mm] 0 \end{array} 
\]
follows the convention of~\eqref{cGZ}, consisting of $j-1$ zero rows at the bottom, and the first $r-j+1$ rows are
of the form $10\cdots0$.
The functions $G_k$ appearing~\eqref{Acj+r}-\eqref{Acj-r} are given by
\begin{align}
G_{k}(\mu_{1r}, \mu_{2r},\ldots, \mu_{rr}) &=
\left(-\frac{
({\cal E}_m(\mu_{kr}+m-n-k)+1)\prod_{j\neq k=1}^{m} (\mu_{kr}-\mu_{jr}-k+j)}
{\prod_{j\neq \frac{k}{2}=1}^{\lfloor m/2 \rfloor} (\mu_{kr}-\mu_{2j,r}-k+2j)
(\mu_{kr}-\mu_{2j,r}-k+2j+1)}
\right)^{1/2} \nn\\
& \times \prod_{j=1}^n\left(\frac{
\mu_{kr}+\mu_{m+j,r}+m-j-k+2}
{\mu_{kr}+\mu_{m+j,r}+m-j-k+2-{\cal E}_{m+\mu_{m+j,r}}}
\right)^{1/2} \label{Gkeven}
\end{align}
for $k\leq m$ and $k$ even; 
\begin{align}
& G_{k}(\mu_{1r}, \mu_{2r},\ldots, \mu_{rr}) =\nn\\
&
\left(\frac{(p-\mu_{kr}+k-1)
({\cal O}_m(\mu_{kr}+m-n-k)+1)\prod_{j\neq k=1}^{m} (\mu_{kr}-\mu_{jr}-k+j)}
{\prod_{j\neq \frac{k+1}{2}=1}^{\lceil m/2 \rceil} (\mu_{kr}-\mu_{2j-1,r}-k+2j-1)
(\mu_{kr}-\mu_{2j-1,r}-k+2j)}
\right)^{1/2} \nn\\
& \times \prod_{j=1}^n\left(\frac{
\mu_{kr}+\mu_{m+j,r}+m-j-k+2}
{\mu_{kr}+\mu_{m+j,r}+m-j-k+2-{\cal O}_{m+\mu_{m+j,r}}}
\right)^{1/2}
\label{Gkodd}
\end{align}
for $k\leq m$ and $k$ odd. The remaining expressions are
\begin{align}
& G_{m+k}(\mu_{1r}, \mu_{2r},\ldots, \mu_{rr}) 
=(-1)^{\mu_{m+k+1,r}+\mu_{m+k+2,r}+\ldots+\mu_{rr}}\nn\\
&
\times \left(
({\cal O}_{\mu_{m+k,r}}(\mu_{m+k,r}-k+n)+1)({\cal E}_{m+\mu_{m+k,r}}(p+\mu_{m+k,r}+m-k)+1)
\right)^{1/2} \nn\\
&
\times \left(\frac{
\prod_{j=1}^{\lfloor m/2 \rfloor} ({\cal E}_{m+\mu_{m+k,r}}(\mu_{2j,r}+\mu_{m+k,r}-2j-k+m+1)+1)}
{\prod_{j=1}^{\lceil m/2 \rceil} ({\cal E}_{m+\mu_{m+k,r}}(\mu_{2j-1,r}+\mu_{m+k,r}-2j-k+m+1)+1)}\right)^{1/2}
 \nn\\
&
\times \left(\frac{
\prod_{j=1}^{\lceil m/2 \rceil} ({\cal O}_{m+\mu_{m+k,r}}(\mu_{2j-1,r}+\mu_{m+k,r}-2j-k+m+2)+1)}
{\prod_{j=1}^{\lfloor m/2 \rfloor} ({\cal O}_{m+\mu_{m+k,r}}(\mu_{2j,r}+\mu_{m+k,r}-2j-k+m)+1)}\right)^{1/2}\nn\\
& \times \prod_{j\neq k=1}^n\left(\frac{
\mu_{m+j,r}-\mu_{m+k,r}-j+k}
{\mu_{m+j,r}-\mu_{m+k,r}-j+k-{\cal O}_{\mu_{m+j,r}-\mu_{m+k,r}}}
\right)^{1/2}
\label{Gm+k}
\end{align}
for $k=1,2,\ldots,n$.

Herein ${\cal E}$ and ${\cal O}$ are the even and odd functions defined by
\begin{align}
& {\cal E}_{j}=1 \hbox{ if } j \hbox{ is even and 0 otherwise},\nn\\
& {\cal O}_{j}=1 \hbox{ if } j \hbox{ is odd and 0 otherwise}. \label{EO}
\end{align}
The notation $\lfloor a \rfloor$ (resp.\ $\lceil a \rceil$)
refers to the {\em floor} (resp.\ {\em ceiling}) of
$a$, i.e.\ the largest integer not exceeding~$a$ (resp.\ the smallest integer greater than or equal to $a$).

The other coefficients appearing in~\eqref{Acj+r}-\eqref{Acj-r} are $\gl(m|n)$ Clebsch-Gordan coefficients 
arising in the tensor product~\eqref{umntensor}.
Such CGCs have been determined in~\cite{CGC}, and can be written as a
product of isoscalar factors in the following way: 
\begin{align}
& 
\left( 
\begin{array}{l}1 0 \cdots 0 0\\[-1mm]
1 0 \cdots 0\\[-1mm] \cdots\\[-1mm] 0 \end{array}  ;
\begin{array}{ll} [\mu]^r \\[2mm] |\mu)^{r-1} \end{array} 
\Bigg| \begin{array}{ll} [\mu]^r_{+k} \\[2mm] |\mu')^{r-1} \end{array} \right)
 =  (-1)^{\sum_{i=1}^m\sum_{q=m}^{j-1}\theta_{iq}}
\left( \begin{array}{l} 1\dot{0} \\ 1 \dot{0} \end{array}
\left| \begin{array}{l}  [\mu]^r \\ {[\mu]}^{r-1} \end{array} \right.\right|
\left. \begin{array}{l} [\mu]^r_{+k}  \\ {[\mu^\prime]}^{r-1} \end{array} \right)\times \ldots \nonumber\\
& \times\left( \begin{array}{l} 1\dot{0} \\ 1 \dot{0} \end{array} 
\left| \begin{array}{l}  [\mu]^{j+1} \\ {[\mu]}^{j} \end{array} \right.\right|
\left. \begin{array}{l} [\mu\prime]^{j+1}  \\ {[\mu^\prime]}^{j} \end{array} \right) 
\left( \begin{array}{l} 1\dot{0} \\ 0 \dot{0} \end{array}
\left| \begin{array}{l}  [\mu]^{j} \\ {[\mu]}^{j-1} \end{array} \right. \right|
\left. \begin{array}{l} [\mu\prime]^{j}  \\ {[\mu^\prime]}^{j-1} \end{array} \right)\times 1.  
 \label{CGC1}
 \end{align} 
As before, the pattern in the left hand side consists of $j-1$ zero rows at the bottom, and the first $r-j+1$ rows are
of the form $10\cdots0$.
In the right hand side of~\eqref{CGC1},
\[
\left( \begin{array}{l} 1 \dot{0} \\ \beta \dot{0} \end{array}
\left|\begin{array}{l} [\mu]^{t} \\ {[\mu]}^{t-1} \end{array} \right.  \right|
\left. \begin{array}{l} [\mu]^{t}_{+k} \\ {[\mu']}^{t-1} \end{array} \right) \qquad (\beta\in\{0,1\})
\]
is a $\gl(m|t-m)\supset\gl(m|t-m-1)$ isoscalar factor when $m+1\leq t\leq m+n$,
and a $\gl(t)\supset\gl(t-1)$ isoscalar factor when $1\leq t\leq m$.
Note that for $j=m+1,\ldots, r$ only $\mathfrak{gl}(m|b)\supset \mathfrak{gl}(m|b-1)$ isoscalar factors appear in~(\ref{CGC1})
and for $j=1,\ldots, m$ both $\mathfrak{gl}(m|b)\supset \mathfrak{gl}(m|b-1)$ and $\mathfrak{gl}(b)\supset \mathfrak{gl}(b-1)$ 
isoscalar factors are present. 

For the $\mathfrak{gl}(m|t-m)\supset \mathfrak{gl}(m|t-m-1)$ isoscalar factors there are six different expressions,
depending on the position of the pattern changes in the right hand side. 
These six expressions are given by:
\begin{align}
&\left( \begin{array}{l} 1 \dot{0} \\0 \dot{0} \end{array}
\left|\begin{array}{l} [\mu]^{t} \\ {[\mu]}^{t-1} \end{array} \right.  \right|
\left. \begin{array}{l} [\mu]^{t}_{+k} \\ {[\mu]}^{t-1} \end{array} \right) 
\nonumber\\
&
= (-1)^{k-1}(-1)^{{\sum_{i=k}^m}\theta_{i,t-1}}
\left( \prod_{i=1(\neq k)}^m \left(\frac{l_{kt}-l_{it}+1}{l_{kt}-l_{i,t-1}}
\right)  
 \frac{\prod_{s=m+1}^{t-1}  
(l_{kt}-l_{s,t-1} )}
{ \prod_{s=m+1}^{t} (l_{kt}-l_{st}+1)}
\right)^{1/2} \quad(1\leq k \leq m);\label{liso1}
\end{align}

\begin{align}
&\left( \begin{array}{l} 1 \dot{0} \\0 \dot{0} \end{array}
\left| \begin{array}{l} [\mu]^{t} \\ {[\mu]}^{t-1} \end{array} \right. \right|
\left. \begin{array}{l} [\mu]^{t}_{+k} \\ {[\mu]}^{t-1} \end{array} \right)
= \left( \prod_{i=1}^m \left(\frac{l_{it}-l_{kt}}{l_{i,t-1}-l_{kt}+1}
\right)  
 \frac{\prod_{s=m+1}^{t-1}  
(l_{s,t-1}-l_{kt}+1 )}
{ \prod_{s=m+1(\neq k)}^{t} (l_{st}-l_{kt})}
\right)^{1/2} \hskip 1.5cm
\nonumber\\& \hskip 8cm  
\quad (m+1\leq k \leq t); 
\label{liso2}
\end{align}

\begin{align}
&\left( \begin{array}{l} 1 \dot{0} \\1 \dot{0} \end{array} 
\left| \begin{array}{l} [\mu]^{t} \\ {[\mu]}^{t-1} \end{array} \right.\right|
\left. \begin{array}{l} [\mu]^{t}_{+k} \\ {[\mu]}^{t-1}_{+q} \end{array} \right)
=(-1)^{k+q}(-1)^{{\sum_{i=\min(k+1,q+1)}^{\max(k-1,q-1)}}\theta_{i,t-1}}S(k,q)  \hskip4cm \nonumber\\
&\times \left( \prod_{i=1(\neq k,q)}^m 
\frac{(l_{i,t-1}-l_{k,t-1}-1-\delta_{kq}+2\theta_{i,t-1})
(l_{i,t-1}-l_{q,t-1})}{(l_{it}-l_{kt})(l_{it}-l_{qt})}
\right)^{\frac{\theta_{q,t-1}}{2}} \nonumber\\  
&
\times\frac{1}{(l_{kt}-l_{qt})^{1-\delta_{kq}}}
\left( \prod_{s=m+1}^{t} \left(
\frac{l_{qt}-l_{st}}
{l_{kt}-l_{st}+1}
\right)
\prod_{s=m+1}^{t-1} \left(
\frac{l_{kt}-l_{s,t-1}}
{l_{q,t-1}-l_{s,t-1}}
\right)
\right)^{\frac{\theta_{q,t-1}}{2}}
\quad (1\leq k,q \leq m);\label{liso3}
\end{align}

\begin{align}
&\left( \begin{array}{l} 1 \dot{0} \\1 \dot{0} \end{array} 
\left| \begin{array}{l} [\mu]^{t} \\ {[\mu]}^{t-1} \end{array} \right. \right|
\left. \begin{array}{l} [\mu]^{t}_{+k} \\ {[\mu]}^{t-1}_{+q} \end{array} \right)
=(-1)^{k}(-1)^{{\sum_{i=1}^{k-1}}\theta_{i,t-1}}
\left(\frac{1}{l_{kt}-l_{q,t-1}}\right)^{1/2}
\nonumber\\
&\times \left( \prod_{i=1(\neq k)}^m \left(\frac{(l_{i,t-1}-l_{k,t-1}-1+2\theta_{i,t-1})
(l_{i,t-1}-l_{q,t-1}+1)}{(l_{it}-l_{kt})(l_{it}-l_{q,t-1})}
\right)\right)^{1/2} \nonumber\\  
& \times
\left( \prod_{s=m+1}^{t} \left(\frac{|l_{st}-l_{q,t-1}|}
{(l_{kt}-l_{st}+1)}
\right) \prod_{s=m+1(\neq q)}^{t-1} \left(\frac{l_{kt}-l_{s,t-1}}
{|l_{s,t-1}-l_{q,t-1}+1|}
\right)\right)^{1/2}\nonumber\\
& \hskip 9cm (1\leq k \leq m, \quad m+1\leq q \leq t-1);\label{liso4}
\end{align}

\begin{align}
&\left( \begin{array}{l} 1 \dot{0} \\1 \dot{0} \end{array} 
\left| \begin{array}{l} [\mu]^{t} \\ {[\mu]}^{t-1} \end{array} \right. \right|
\left. \begin{array}{l} [\mu]^{t}_{+k} \\ {[\mu]}^{t-1}_{+q} \end{array} \right)
=(-1)^{q}(-1)^{{\sum_{i=q+1}^{m}}\theta_{i,t-1}}
\left(\frac{1}{l_{qt}-l_{kt}+1}\right)^{1/2}
\nonumber\\
&\times \left( \prod_{i=1}^m \left(\frac{l_{it}-l_{kt}}{l_{i,t-1}-l_{kt}+1}
\right) \prod_{i=1(\neq q)}^m \Big|\frac{l_{q,t-1}-l_{i,t-1}}{l_{qt}-l_{it}}
 \Big| 
 \prod_{s=m+1(\neq k)}^{t} \Big|\frac{l_{qt}-l_{st}}{l_{st}-l_{kt}}
\Big|
 \prod_{s=m+1}^{t-1} \Big|\frac{l_{s,t-1}-l_{kt}+1}{l_{qt}-l_{s,t-1}-1}
 \Big|\right)^{1/2}\nonumber\\
 & \hskip 9cm (m+1\leq k \leq t, \quad 1\leq q \leq m);\label{liso5}
\end{align}

\begin{align} 
&\left( \begin{array}{l} 1 \dot{0} \\1 \dot{0} \end{array} 
\left| \begin{array}{l} [\mu]^{t} \\ {[\mu]}^{t-1} \end{array} \right.\right|
\left. \begin{array}{l} [\mu]^{t}_{+k} \\ {[\mu]}^{t-1}_{+q} \end{array} \right)
=S(k,q) (-1)^{{\sum_{i=1}^m}\theta_{i,t-1}}\left( \prod_{i=1}^m \left(\frac{(l_{it}-l_{kt})
(l_{i,t-1}-l_{q,t-1}+1)}{(l_{i,t-1}-l_{kt}+1)(l_{it}-l_{q,t-1})}
\right)\right)^{1/2} \nonumber\\ 
& \times
\left( \prod_{s=m+1(\neq k)}^{t} \Big|\frac{l_{st}-l_{q,t-1}}
{l_{st}-l_{kt}}\Big|
 \prod_{s=m+1(\neq q)}^{t-1} \Big| \frac{l_{s,t-1}-l_{kt}+1}
{l_{s,t-1}-l_{q,t-1}+1}
\Big| \right)^{1/2} 
\quad(m+1\leq k \leq t, \quad m+1\leq q \leq t-1).\label{liso6}
\end{align} 

For the $\mathfrak{gl}(t)\supset \mathfrak{gl}(t-1)$ isoscalar factors~\cite{Vilenkin,paraboson} there are only two 
different expressions, given by:
\begin{equation}
\left( \begin{array}{l} 1 \dot{0} \\0 \dot{0} \end{array} 
\left| \begin{array}{l} [\mu]^t \\ {[\mu]}^{t-1} \end{array} \right. \right|
\left. \begin{array}{l} [\mu]^t_{+k} \\ {[\mu]}^{t-1} \end{array} \right)
= \left( \frac{\prod_{i=1}^{t-1}  
(l_{i,t-1}-l_{kt}-1 )}
{ \prod_{i=1(\neq k)}^{t} (l_{it}-l_{kt})}
\right)^{1/2} 
\label{iso1LA}
\end{equation}   

\begin{equation}
\left( \begin{array}{l} 1 \dot{0} \\ 1 \dot{0} \end{array} 
\left| \begin{array}{l} [\mu]^t \\ {[\mu]}^{t-1} \end{array} \right.\right|
\left. \begin{array}{l} [\mu]^t_{+k} \\ {[\mu]}^{t-1}_{+q} \end{array} \right)
= S(k,q)\left( \frac{{\prod_{i=1(\neq q)}^{t-1}}  
(l_{i,t-1}-l_{kt}-1 )\prod_{i=1(\neq k)}^{t}(l_{it}-l_{q,t-1})}
{ \prod_{i=1(\neq k)}^{t} (l_{it}-l_{kt})\prod_{i=1(\neq q)}^{t-1}(l_{i,t-1}
 -l_{q,t-1}-1)}
\right)^{1/2}.
\label{iso2LA}
\end{equation}
In all of the formulas above, $\dot{0}$ stands for an appropriate sequence of zeros in the pattern, and we use   
\begin{align}
& l_{is}=\mu_{is}-i+m+1\qquad (1\leq i \leq m),\nn\\
& l_{is}=-\mu_{ps}+i-m \qquad(m+1\leq i\leq s),\nn\\
& S(k,q) = \left\{ \begin{array}{rcl}
 {1} & \hbox{for} & k\leq q  \\ 
 {-1} & \hbox{for} & k>q.
 \end{array}\right. ,
 \label{S}
\end{align} 
and the obvious convention that $\prod_{i=a(\neq k)}^{b}$ means ``the product over all $i$-values
running from $a$ to $b$, but excluding $i=k$.

\section*{Acknowledgments}
The authors were supported by the Joint Research Project ``Lie superalgebras -- applications in quantum theory'' in the framework of an international collaboration programme between the Research Foundation -- Flanders (FWO) and the Bulgarian Academy of Sciences. NIS was partially supported by Bulgarian NSF grant DFNI T02/6.


\begin{thebibliography}{99}

\bibitem{Green}
Green H S 1953 
A Generalized Method of Field Quantization
{\em Phys.\ Rev.} {\bf 90} 270-273 

\bibitem{Kamefuchi}
Kamefuchi S and Takahashi Y 1962
A generalization of field quantization and statistics
{\em Nucl.\ Phys.} {\bf 36} 177-206 

\bibitem{Ryan}
Ryan C and Sudarshan E C G 1963 
Representations of parafermi rings
{\em Nucl.\ Phys.} {\bf 47} 207-211 

\bibitem{Ganchev}
Ganchev A Ch and Palev T D 1980 
A Lie Superalgebraic Interpretation of the Para-Bose Statistics
{\em J.\ Math.\ Phys.} {\bf 21} 797-799 

\bibitem{GM}
Greenberg O W and Messiah A M 1965 
Selection Rules for Parafields and the Absence of Para Particles in Nature 
{\em Phys.\ Rev.} {\bf 138} (5B) 1155-1167 

\bibitem{Palev1}
Palev T D 1982
Para-Bose and para-Fermi operators as generators of orthosymplectic Lie superalgebras
{\em J.\ Math.\ Phys.} {\bf 23} 1100-1102 
 
\bibitem{YJ}
Yang W and Jing S 2001 
A New Kind of Graded Lie Algebra and Parastatistical Supersymmetry
{\em Science in China (Series A)} {\bf 44} 1167-1173

\bibitem{YJ2}
Yang W and Jing S 2001 
Fock Space Structure for the Simplest Parasupersymmetric System
{\em Mod.\ Phys.\ Lett.\ A} {\bf 16} 963-971

\bibitem{KA}
Kanakoglou K and Herrera-Aguilar A 2011 
Graded Fock-like representations for a system of algebraically interacting paraparticles 
{\em J.\ Phys.: Conf.\ Series} {\bf 287} 011237

\bibitem{KK}
Kanakoglou K 2011 
Ladder Operators, Fock-Spaces, Irreducibility and Group Gradings for the Relative Parabose Set Algebra
{\em Int.\ J.\ Alg.} {\bf 5} 413-428
 
\bibitem{Tolstoy2014}
Tolstoy V N 2014
Once more on Parastatistics
{\em Phys.\ Part.\ Nucl.\ Lett.} {\bf 11} 933-937

\bibitem{parafermion}
Stoilova N I and Van der Jeugt J 2008 
The parafermion Fock space and explicit $\mathfrak{so}(2n+1)$ representations
{\em J.\ Phys.\ A: Math.\ Theor.} {\bf 41} 075202 

\bibitem{paraboson}
Lievens S, Stoilova N I and Van der Jeugt J 2008 
The paraboson Fock space and unitary irreducible representations of the Lie superalgebra $\mathfrak{osp}(1|2n)$
{\em Commun.\ Math.\ Phys.} {\bf 281} 805-826 


\bibitem{SV2015}
Stoilova N I and Van der Jeugt J 2015 
A class of infinite-dimensional represenations of the Lie superalgebra $\mathfrak{osp}(2m+1|2n)$ and the parastatistics Fock space
{\em J.\ Phys.\ A: Math.\ Theor.} {\bf 48} 155202 

\bibitem{CGC}
Stoilova N I and Van der Jeugt J 2010 Gel'fand-Zetlin basis and Clebsch-Gordan coefficients for covariant representations of the Lie superalgebra $\gl(m|n)$ 
{\em J.\ Math.\ Phys.} {\bf 51} 093523 

\bibitem{Werry}
Werry J L, Isaac P S and Gould M D 2017
Reduced Wigner coefficients for Lie superalgebra $gl(m|n)$ corresponding to unitary representations and beyond
{\em J.\ Phys.\ A: Math.\ Theor.} {\bf 50} 365202

\bibitem{Rittenberg1}
Rittenberg V and Wyler D 1978
Generalized Superalgebras
{\em Nucl.\ Phys.\ B} {\bf 139} 189-202

\bibitem{Rittenberg2}
Rittenberg V and Wyler D 1978
Sequences of $\Z_2\times \Z_2$-graded Lie algebras and superalgebras
{\em J.\ Math.\ Phys.} {\bf 19} 2193-2200

\bibitem{Aizawa1}
Aizawa N, Kuznetsova Z, Tanaka H and Toppan F 2016
$\Z_2\times\Z_2$-graded Lie symmetries of the L\'evy-Leblond equations
{\em Prog.\ Theor.\ Exp.\ Phys.} {\bf 2016} 123A01

\bibitem{Aizawa2}
Aizawa N and Segar J 2017
$\Z_2\times\Z_2$ generalizations of $N = 2$ super Schr\"odinger algebras and their representations
(arXiv:1705.10414) 

\bibitem{Kac}
Kac V G 1977 
Lie superalgebras 
{\em Adv.\ Math.} {\bf 26} 8-96 

\bibitem{Kac1}
Kac V G 1978 
Representations of Classical Lie Superalgebras 
{\em Lect.\ Notes in Math.} {\bf 626} 597-626 



\bibitem{Mac}
Macdonald I G 1995 
{\em Symmetric Functions and Hall Polynomials}, 2nd edition  (Oxford: Oxford University Press)

\bibitem{Berele}
Berele A and Regev A 1987
Hook Young diagrams with applications to combinatorics and to representations of Lie superalgebras
{\em Adv.\ Math.} {\bf 64} 118-175


\bibitem{King1990}
King R C 1990
S-functions and characters of Lie algebras and superalgebras
{\em IMA Volumes in Mathematics and its Applications} {\bf 19} 226-261 

\bibitem{Vilenkin}
Vilenkin N J and A.U. Klimyk A U 1992
{\em Representation of Lie Groups and Special Functions}, Vol.~3
(Amsterdam: Kluwer Academic Publishers)


\end{thebibliography}
\end{document}